Original Paper

# The Impact of Explanations on Layperson Trust in Artificial Intelligence–Driven Symptom Checker Apps: Experimental Study

Claire Woodcock[1], BA, MSc; Brent Mittelstadt[1], BA, MA, PhD; Dan Busbridge, MPhys, PhD; Grant Blank[1], BA, MA, PhD

[1]Oxford Internet Institute, University of Oxford, Oxford, United Kingdom

**Corresponding Author:**
Claire Woodcock, BA, MSc
Oxford Internet Institute
University of Oxford
1 St Giles
Oxford, OX1 3JS
United Kingdom
Phone: 44 1865 287210
Email: cwoodcock.academic@gmail.com

## Abstract

**Background:** Artificial intelligence (AI)–driven symptom checkers are available to millions of users globally and are advocated as a tool to deliver health care more efficiently. To achieve the promoted benefits of a symptom checker, laypeople must trust and subsequently follow its instructions. In AI, explanations are seen as a tool to communicate the rationale behind black-box decisions to encourage trust and adoption. However, the effectiveness of the types of explanations used in AI-driven symptom checkers has not yet been studied. Explanations can follow many forms, including *why*-explanations and *how*-explanations. Social theories suggest that *why*-explanations are better at communicating knowledge and cultivating trust among laypeople.

**Objective:** The aim of this study is to ascertain whether explanations provided by a symptom checker affect explanatory trust among laypeople and whether this trust is impacted by their existing knowledge of disease.

**Methods:** A cross-sectional survey of 750 healthy participants was conducted. The participants were shown a video of a chatbot simulation that resulted in the diagnosis of either a migraine or temporal arteritis, chosen for their differing levels of epidemiological prevalence. These diagnoses were accompanied by one of four types of explanations. Each explanation type was selected either because of its current use in symptom checkers or because it was informed by theories of contrastive explanation. Exploratory factor analysis of participants' responses followed by comparison-of-means tests were used to evaluate group differences in trust.

**Results:** Depending on the treatment group, two or three variables were generated, reflecting the prior knowledge and subsequent mental model that the participants held. When varying explanation type by disease, migraine was found to be nonsignificant ($P=.65$) and temporal arteritis, marginally significant ($P=.09$). Varying disease by explanation type resulted in statistical significance for input influence ($P=.001$), social proof ($P=.049$), and no explanation ($P=.006$), with counterfactual explanation ($P=.053$). The results suggest that trust in explanations is significantly affected by the disease being explained. When laypeople have existing knowledge of a disease, explanations have little impact on trust. Where the need for information is greater, different explanation types engender significantly different levels of trust. These results indicate that to be successful, symptom checkers need to tailor explanations to each user's specific question and discount the diseases that they may also be aware of.

**Conclusions:** System builders developing explanations for symptom-checking apps should consider the recipient's knowledge of a disease and tailor explanations to each user's specific need. Effort should be placed on generating explanations that are personalized to each user of a symptom checker to fully discount the diseases that they may be aware of and to close their information gap.

*(J Med Internet Res 2021;23(11):e29386)*   doi: 10.2196/29386

# Introduction

## Overview

Health care is a need so universal that the right to adequate medical care is enshrined in the Universal Declaration of Human Rights [1]. Yet, globally, governments face long-term challenges. High-income countries struggle with the financial burden of providing health care to aging populations with complex needs [2]. Meanwhile, half the world's population "lacks access to essential health services [3]..." More immediately, the COVID-19 pandemic is straining health care systems while making it necessary for care to be delivered remotely where possible [4].

These pressures have cultivated interest in "tools that use computer algorithms to help patients with self diagnosis or self triage," called symptom checkers (SCs) [5]. Although SCs have been developed using a broad range of techniques including Bayesian [6], rule-based, and deep learning methods [7], they are generally referred to as using artificial intelligence (AI) [8]. SCs are typically presented as smartphone chatbot apps and are created for one of two aims. First, akin to a visit to a primary care practitioner, some SCs are built to allow individuals to check what may be causing their health symptoms out of all common health conditions. Second, as a specific subset, some SCs check for symptoms of one disease only, typically COVID-19.

Private companies who build these apps [8] and government officials [9] believe SCs can improve provision of health care in two ways: (1) those with benign conditions can be easily triaged to less resource-intensive care, allowing human clinicians to focus on patients in need [10], and (2) SCs reduce the need for individuals to travel. This drives efficiencies in high-income countries [11] and allows those in countries with less care provision to access medical advice from remote locations.

Given their rapid global deployment, SCs have faced much scrutiny. Debate has focused on their accuracy, including their potential to fail to detect dangerous illnesses [12] or to provide overly cautious diagnoses [5,8]. Equally as important is the *human aspect* of SCs. By their nature, SCs abstract from the human interaction of a patient and physician [5]. The presented diagnosis may or may not provide additional insights as to why the app has come to a particular decision (Multimedia Appendix 1). Even assuming that an SC provides an accurate diagnosis, lay individuals must still follow triage instructions to achieve the intended benefits of SCs at a population level. Ensuring that users trust SCs is therefore of paramount importance if SCs are to achieve their intended benefits with regard to reducing the pressures faced by health care systems globally.

As with SCs, trustworthiness is seen as a necessary requirement for widespread adoption of AI. An essential component of trustworthiness is the capacity of a system (or its operator) to explain its behavior, for example, the rationale behind a particular diagnosis. Explanations are seen as a tool to communicate the rationale behind black box decisions to encourage user trust and adoption.

Recent studies have started to examine the effectiveness of different types of AI explanations [13-15], but to date no studies have specifically looked at SC explanations. Qualitative analysis of general sentiment toward medical conversational AI agents reveals a mixed reception [16], which suggests that choosing the right type of explanation in SCs is critically important for the systems to be well received. The lack of study of explanations in SCs poses challenges because poor explanations could reduce a person's inclination to use SCs, fuel health anxiety [17], or cause them to seek a second opinion from a human clinician, all of which would further burden health care systems. Furthermore, explanations can take many forms, including *why*-explanations and *how*-explanations, each of which may encourage user trust to different degrees. Social theories suggest that *why*-explanations are better at communicating knowledge and cultivating trust in laypeople, but this hypothesis has yet to be tested for SCs.

This paper presents the results of an exploratory study of layperson perception of SC explanations. Trust is used as a measurement of explanatory quality because good explanations are known to increase trust in AI systems [18], increasing the likelihood that the recipient will follow its output [19]. In this section, we begin by grounding the study in philosophical theories of contrastive explanation alongside cognitive psychology studies of causality. These theories emphasize that humans require explanations when information gaps are created. Although SCs currently address the need for explanation by explaining how the system derived the answer, humans typically prefer *why*-explanations. In the *Methods* section, we discuss the methodology used to conduct a study of 750 laypeople, where each participant was presented with a diagnosis of one of two diseases that was accompanied by one of four explanations. The results are presented in the *Results* section. In the *Discussion* section, we discuss findings that suggest trust may vary by explanation type for a lesser-known disease and that trust in explanation is significantly affected by the disease being explained. In the *Conclusions* section, we provide recommendations for SC system builders. The data and code required to reproduce all findings are publicly available [20].

## Explanations in Theory and Practice

### Overview

As this study focuses on the impact of varying types of SC explanations on layperson trust, it is first necessary to understand the purpose and use of explanations in AI. Explanations in theory and practice have long been studied by researchers, resulting in expansive literature on the subject. Here we adopt Lewis' [21] definition of an explanation as the provision of the causal history of an event. We also draw on Hilton [22] and Miller [23] to argue that explanations are ambiguous in being both a verb and a noun where an explainer *explains* (verb) *something* (noun) to *generate understanding* (verb) in a recipient.

### Explanations and Their Purpose

Explanatory theory has typically focused on the function of explanations as a mechanism to transmit information about the *explanandum* (the event or phenomenon being explained) to further inform a recipient's knowledge [24]. However, explanations can be used in many ways, including to persuade [25], assign blame [26], or even deceive the recipient [27,28]. Given these purposes, the explainer and the recipient may frequently have differing goals [25].

The goal of the SC explanation is to generate enough trust in the recipient so that they will follow triage instructions to reduce health system burden [8,10]. This is distinct from the goal of the recipient, which is to understand what is causing their symptoms [29].

*Explanation Seeking and Knowledge*

Explanations may be prevalent in human and social interactions, but they are not ubiquitous. People select when to seek an explanation [26]. Intrinsic provocation to seek an explanation, referred to as explanation seeking curiosity, is strongly predicted by future learning and future utility and moderately predicted by lack of knowledge [30].

The influential information gap theory described by Lowenstein [31] proposes that a delta between an individual's current knowledge and their desired knowledge cultivates curiosity. It is normatively correct to seek an explanation when an event or phenomenon does not fit one's mental model [32]. Empirically, information gaps have been shown to provoke explanation seeking in both children [33] and adults [34].

Knowledge and explanation are intimately linked [35]. When presented with information, the recipient must be aware of their lack of knowledge to seek an explanation [31]. An explanation must sufficiently transmit information [36], which is evaluated against prior beliefs and knowledge, particularly in knowledge-rich domains [25], the desired result being an updated mental model in the recipient [26]. The recipient's perception of an explanation subsequently provides insight into their knowledge of the phenomenon or the event being explained [22,37]. Hence, explanatory quality is critical because "explanations that are good and are satisfying to users enable users to develop a good mental model. In turn, their good mental model will enable them to develop appropriate trust in the AI [18]."

*Everyday Explanations*

Explanations can take many forms, for example, scientific, causal, teleological, or *everyday*. A universally accepted taxonomy does not exist [38]. Given this study's focus on laypeople and prior scholarship suggesting their utility in AI [23,38], our focus will be on everyday explanations. These are a form of explanation, which are commonly observed in social interaction, defined as an answer to a *why* question [21,23,39].

Lewis [21] asserts that most explanations are typically answers to *why* questions, for example, "Why did you do that?" or "Why did that happen?" *Why* questions are implicitly contrastive, with the form of the question being "why that decision *rather than something else?* [21]." When humans answer a *why* question, we offer an explanation (P) relative to some other event that did not occur (Q). This is termed a contrastive explanation, where P is referred to as the *fact* (an event that occurred) and Q is the *foil* (an event that did not occur) [39].

The factual component of a *why* question can have many potential foils. Consider the question "Why did you watch Game of Thrones?" Its foils could include "rather than the news?" or "instead of going out?" The foil itself generates context for the explanation to be provided. In human interactions, foils are often not explicitly stated in the *why* question. Instead, humans infer the foil from the tone and context of the interaction [39]. Importantly, the cause explained is dependent on the questioner's interest implied through conversation. This further emphasizes the assertion by Miller [23] that explanations are social and conversational [22].

Explaining the contrast can be easier than explaining the fact itself because P does not need to be sufficient for the event to occur, provided it differentiates between the causal difference of P and Q [40]. Contrastive explanations also have the benefit of constraining the information to be provided [39]. The constraining effect of a contrastive explanation is helpful to humans because it reduces the cognitive burden of processing an explanation [26].

Humans rarely provide an exhaustive causal chain as an explanation, preferring instead to select one or two pertinent causes [25,36]. In the example of the television show choice, a layperson would not justify the selection by providing their life story, listing influential childhood events. Instead, the explainer might answer, "It's more exciting than real life," and "I was tired," respectively. These explanations may not entirely explain P; however, they sufficiently and succinctly differentiate between P and Q. For the remainder of this paper, for simplicity, contrastive explanations will be referred to as *why*-explanations.

*Explanation Complexity*

Striking an appropriate balance in terms of the complexity of an explanation is difficult. Thagard's [41] theory of explanatory coherence states that people prefer simple, general explanations. This preference has been validated empirically [23]. For example, Read and Marcus-Newhall [42] evaluated this using a scenario about a woman with three symptoms: weight gain, fatigue, and nausea. The study participants received one of the following three explanation types describing what was causing her ill health:

1. Narrow:
    a. Having stopped exercising (explains weight gain).
    b. Having mononucleosis (explains fatigue).
    c. Having a stomach virus (explains nausea).
2. Broad: she is pregnant (explaining all three).
3. Conjunctive: all of the causes in 1 are true.

The participants preferred the broad explanation of pregnancy (option 2), preferring simple explanations that had fewer causes and explained more events.

Contemporary laboratory-based studies reiterate human preference for simple explanations over complex ones [13,36,43-45]. However, experiments in natural settings have revealed that complexity increases explanatory satisfaction [28,46] and that complexity preferences are aligned with the complexity of the event itself [47-49].

Frequently, studies in cognitive psychology examine diagnostic explanation through discussion of an alien race's illnesses to avoid reliance on prior knowledge. This removes a confounder; however, we should be cautious applying the findings to real-life SCs, given the relationship between explanation and knowledge (see *Explanation Seeking and Knowledge* section). Importantly, most of the experiments required high levels of literacy and comprehension. They were performed on cognitive psychology students or recruits screened for literacy ability. This is in contrast with reality: 1 in 7 of the UK population are functionally illiterate and would struggle to read a medicine label [50], and only half have an undergraduate degree [51].

In short, variance in explanatory preference has been noted between laboratory and natural experimental settings. In addition, experiments have been conducted on study groups with different characteristics to the general population. This reveals a need to validate preferences before generalizing to technologies used by the layperson population.

## Explanations in AI

### Overview

It should not be assumed that the presentation of explanations in AI systems matches with human explanatory behaviors and needs. Whereas the need for humans to explain themselves is often taken for granted in particular situations, there is continued debate around whether it is necessary for AIs to explain themselves. Turing Award winner Geoffrey Hinton argues that explanations are not necessary because humans cannot explain their own neural processes [52]. A study of medical students supports this because half of them were found to rely on intuitive thinking in diagnostic decision-making [53].

Recognizing that AI systems are frequently used to (help) make impactful decisions, explanations can be said to be required for at least two reasons. First, explanations are necessary for users to adopt AI technologies. When explanations are provided, trust and propensity to rely on systems increases [23,54-57]. If AI systems are not trusted, they are less likely to be adopted by their users, limiting the efficacy of the technology [19,58]. Second, the General Data Protection Regulation, the European Union law for data protection and privacy, places two relevant requirements on AI systems: they (1) must provide "meaningful information about the logic involved in the decision-making process" and (2) should ideally provide "an explanation of the decision [59]."

This results in a focus on explaining the decision-making of AI systems [57]. To do this, the algorithmic processes are examined to generate an explanation (the product), and the contained knowledge is subsequently communicated to the recipient (a social process) [23]. In contrast to human explanatory preferences, current AI explanations predominantly provide answers to *how* questions, not *why* questions.

### How-Explanations in AI

AI models are often treated as black boxes due to their complexity and opacity. The explainable AI field predominantly focuses on increasing the transparency of how models produce their outputs, typically for use by computer programmers and expert users. This essentially answers a *how* question: "How did you decide that?"

There are a vast number of techniques that are used to explain how an AI model comes to a decision or otherwise produces an output [13]. These functional explanations will be referred to as *how*-explanations. Two such *how*-explanations are currently used to explain the outputs of SCs (see *Explanations in Symptom Checkers* section):

1. Input influence: Presents a list of the variables input to the model with a quantitative measure of their contribution (positive or negative) to the outcome [60]. Consider a system that reads mammograms. An input influence explanation might highlight visual areas of the scan that strongly influenced its diagnosis of a tumor.
2. Case-based reasoning: Displays a case from the model's training data that is closest to the one being classified [61]. In the mammogram example, a case-based reasoning explanation might declare the scan to be negative and provide another individual's mammogram to explain the verdict.

Both these explanatory methods require their human recipient to have domain knowledge to evaluate the explanation. Neither provides an explicitly contrastive explanation or an answer to a *why* question. Instead, a cognitive burden is placed on a human expert recipient to evaluate and contrast a large number of data points to determine whether they agree with the decision. In the mammogram example, a person not trained in radiology would have insufficient knowledge to understand either explanation, potentially resulting in the recipient perceiving them as poor quality.

When AI systems are deployed to wider society, system builders (such as software engineers and designers) must package the technology into software. For general audiences, the system builders take these *how*-explanations and translate them into a form comprehensible to a nonexpert [15]. Although this approach seems natural to many computer scientists and lawyers, it is not necessarily the best approach for SCs.

SCs, like many AI systems, are often presented as conversational agents or assistants. Humans are prone to anthropomorphism [62], and virtual assistants have many features that lead users to infer human-like agency in the assistants' behavior [63]. In human conversation, we prefer social *why*-explanations (see *Everyday Explanations* section). If SCs are to act as conversational agents, this suggests that the explanations being offered should adhere to norms of human conversation by creating a shared understanding of the decision made between the system and a human recipient [64]. Given that it is system builders who currently generate explanations for AI systems, the large body of work on explanations rooted in sociological, philosophical, and cognitive theories is underused [38].

### Why-Explanations in AI

The conflict between layperson perceptions of *how*-explanations and *why*-explanations is the focus of this study. Given human preferences, the hypothesis is that there will be a preference for *why*-explanations of SC outputs. As such, a type of

XSL•FO
RenderX

*why*-explanation that has been proposed as more effective and accessible will be included, the counterfactual explanation.

Counterfactual explanations present how the factors considered to reach a decision must change for an alternative decision to be made [26,40,65]. A counterfactual explanation implicitly answers the *why* question "Why did you decide outcome P rather than outcome Q?" by examining what would happen if the variables $V = (v_1, v_2,...)$ were different [65,66]. Returning to the mammogram example, a user may ask "*Why* did you diagnose a tumor [rather than no tumor]?" to which the counterfactual explanation may state "If *these* pixels were not white, I would not have diagnosed a tumor."

Counterfactual explanations appear naturally in human cognition. They are a pattern that feature prominently in our day-to-day thoughts [67], with the capability to think counterfactually emerging around the age of two [68]. They are also contrastive, aligning with human explanatory preferences (see *Everyday Explanations* section). Counterfactual explanations are considered an efficient way of communicating causal reasoning [66,69]. They are effective in highlighting model discrimination [70] and, importantly for this work, they offer a type of *why*-explanation [27,40,66]. Counterfactual explanations do require cognitive effort on the recipient's part. However, by their nature, they bound the scope of explanation, reducing cognitive burden [69], and are advocated as a more accessible method for nontechnical users [65].

This study aims to investigate layperson trust in SC explanations. Humans have strong preferences for *why*-explanations, whereas technical methods developed to explain AI systems to date typically give *how*-explanations. Three explanation types of relevance have been identified for evaluation. To directly address this study's focus, the current state of explanations in SCs were examined. SCs, like other AI systems, currently tend to provide *how*-explanations.

### Explanations in Symptom Checkers

To explore the current use of explanations in SCs, we surveyed 10 commercially available SCs (Multimedia Appendix 1). All SCs surveyed were presented in chatbot format to provide a natural mechanism for data collection from the user, mimicking their experience of speaking to a clinician. This further reinforces the view that "causal explanation takes the form of conversation and is thus subject to the rules of explanation [22]," that is, the recipient foremost requires a *why*-explanation.

Current SCs do not allow users to indicate what kind of explanation they require. Instead, the explanation type and content are predefined by system builders. The explanations provided are succinct, typically consisting of a single sentence. This succinctness is likely driven by software user-experience principles, which classify complexity as a detractor to technology adoption [71]. Again, this presentation contrasts with the rich explanations typically generated in the explainable AI field for expert users.

The SCs typically presented a form of explanation alongside the suggested diseases causing symptoms. They were observed taking two forms: (1) a contraction of an input influence explanation which provided one or two health symptoms that most positively influenced the SC's decision; (2) a social proof.

Social proofs, popularized to system builders by Cialdini [72] and Eyal [73] are based on extensive psychological studies that demonstrate that humans are both consciously and unconsciously susceptible to others' cues when making decisions. Social proof tactics can include cues such as user reviews and likes on social media. In the case of SCs, social proof is offered by explaining how many people with the same symptoms have previously been diagnosed with a particular disease. Generating a more detailed view of a social proof would involve providing details of other cases classified by the model, that is, a case-based explanation. Consequently, a social proof explanation can be viewed as a contraction of a case-based reasoning explanation.

This evaluation of current SCs shows that they either provide an input influence explanation or a social proof explanation, both of which are *how*-explanations.

Against this backdrop of the purpose of explanations, human explanatory preferences, and explanation use in SCs, this exploratory study seeks to answer 2 research questions:

Research question 1: Does the type of explanation affect a layperson's trust in the explanation provided by a symptom checker?

Research question 2: Is the person's level of trust affected by existing knowledge of a disease?

## Methods

### Experimental Design

To answer these questions, a 2×4 between-subjects experimental design was constructed, with the participants randomly assigned to a treatment group.

To answer research question 1, four explanation types were selected that reflect the state of the art in explanations of AI and SC outputs: input influence, social proof, counterfactual explanation, and no explanation. In the case of the no explanation type, no specific statement was presented that alluded to how or why the model came to a decision; it was included to provide a baseline for the perception of explanatory quality in chatbot interactions. In the mammogram example, the no explanation type would simply output, "This scan is negative."

To address research question 2, two diseases were selected that had differing levels of population awareness:

1. Migraine: This was chosen as the well-known disease because it affects 1 in 7 of the population [74], making it a disease with high population awareness whose symptoms are widely understood.
2. Temporal arteritis: This was selected as the lesser-known disease because it has a low incidence rate of approximately 0.035% in individuals aged above 50 years [75].

Both diseases involve head pain, which was selected for relatability as the majority of the population have experienced headaches [76]. Epidemiological data were used as a proxy for layperson knowledge to limit study scope because knowledge, like trust, is an intangible variable to measure. The explanations shown are presented in Table 1. Analysis of variance (ANOVA) was planned to compare mean scores across the treatment groups. Given the limited prior research, the anticipated effect size was unknown, and a small effect size (Cohen's $d$=0.2) was chosen. G*Power (Heinrich Heine University) was used to calculate the sample size required for this effect size among 4 groups with a power $\beta$=.8. Erring on the side of caution, a goal of 200 participants per group was set.

**Table 1.** Explanations shown to participants for each combination of the four explanation types and two diseases.

| Explanation type and Disease | Explanation |
| --- | --- |
| **Input influence** | |
| Migraine | "I think this because you have a headache that came on slowly and you feel nauseous." |
| Temporal arteritis | "I think this because you have severe head pain in the side of your temples and your scalp and jaw are painful." |
| **Social proof** | |
| Migraine | "I think this because 8,217 people with your symptoms were diagnosed with a Migraine." |
| Temporal arteritis | "I think this because 8,217 people with your symptoms were diagnosed with Temporal Arteritis." |
| **Counterfactual explanation** | |
| Migraine | "If you didn't feel sick I'd have suggested you have a tension headache." |
| Temporal arteritis | "If your scalp and jaw didn't hurt, I'd have suggested you had a Migraine." |
| **No explanation** | |
| Migraine | No statement presented |
| Temporal arteritis | No statement presented |

## Stimulus Design

Chatbot simulation videos were created using a design and prototyping tool (Botsociety Inc). Information was presented in the manner of the Calgary-Cambridge model of physician-patient consultation [77], in line with the conversational presentation style used in current SCs. To avoid visual elements from confounding the experiment, the chatbot design was limited to text interaction. This contrasts with modern SCs, which use graphics to indicate additional explanatory factors such as the AI-model confidence.

To verify stimuli presentation and question phrasing (see *Creating a Trust Measurement Scale* section), the stimuli and survey were piloted by conducting cognitive interviews with 11 individuals ranging in age from 28 to 62 years. The interviews revealed that at the end of the SC interaction, the participants expected to receive information that matched their typical consultation experiences. This included information about the disease and medical safety nets. Safety nets are a clinical management strategy to ensure the monitoring of patients who are symptomatic, with the aim of avoiding a misdiagnosis or nontreatment of a serious disease, for example, "If your symptoms persist or get worse, please seek medical advice." Consequently, these were included. The final SC videos were approximately 3 minutes in length, with most of the content being identical for each disease. Once the SC came to a conclusion, information was presented in the order shown in Figure 1.

**Figure 1.** The order of information presented at the end of the symptom checker flow.

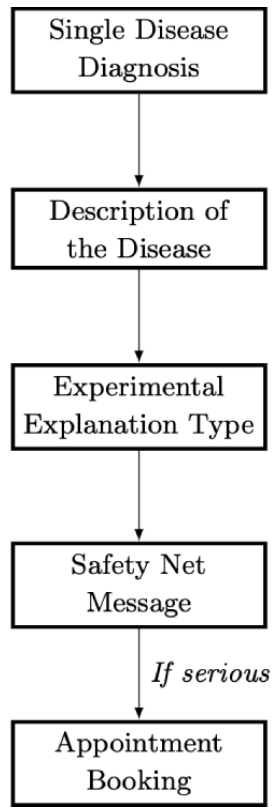

The content of the explanations provided was set according to a medical advisor's clinical knowledge. For input influence, the two symptoms most likely to indicate the disease to a doctor were selected. Likewise, with counterfactual explanation, the symptom most likely to change a clinician's resulting opinion was chosen. For social proof, it was decided not to mirror current SC presentations (eg, "8/10 people with X had symptom Y") because it is known that probability information affects perceptions of explanations [36,44]. Instead, it was opted to state a large number of reference cases, as cognitive interviews indicated that lower numbers suggested a less-advanced AI model. Example screenshots of the stimuli are presented in Figure 2.

**Figure 2.** Screenshots of chatbot stimulus showing the start (left) and end (right) of social proof treatment of migraine.

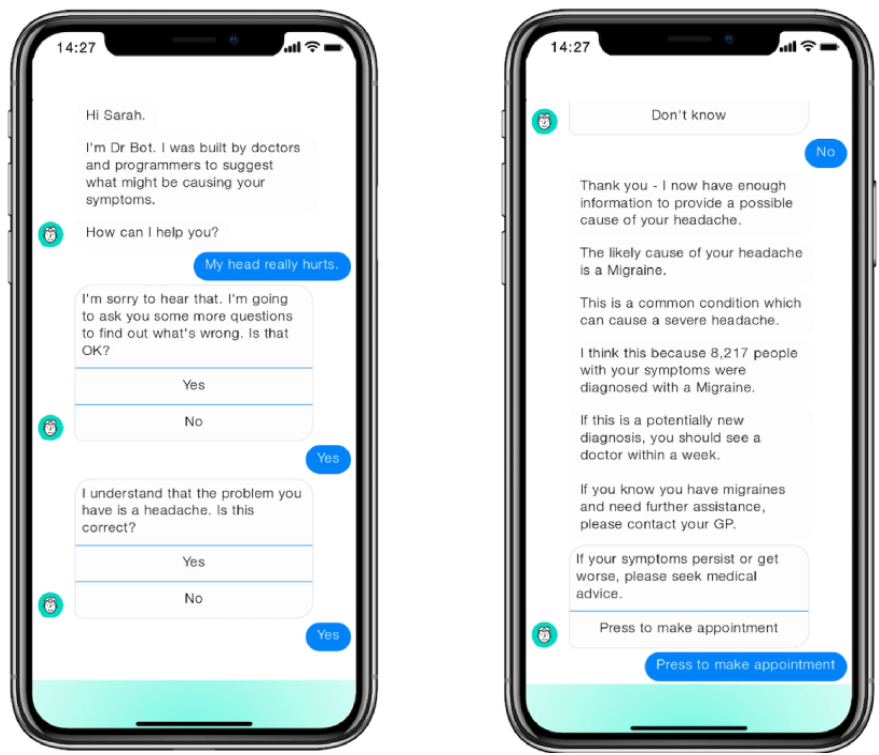

## Creating a Trust Measurement Scale

Trust is a hypothetical construct that cannot be observed or measured in a direct, physical sense [78]. Measurement scales for AI systems must minimally investigate two questions: "Do you trust the output?" (faith) and "Would you follow the system's advice?" (reliance) [79]. Experiments that have analyzed the reliability of trust scales have found high Cronbach's α values, suggesting that these scales are a reliable tool for measuring trust [80]. Nevertheless, there is a lack of agreement in the human-computer interaction field of how to measure trust in explanations. Hoffman [18] highlights that many scales are specific to the application context. Existing scales are also oriented toward evaluating opinions from expert users who repeatedly use a system over periods of time. This is in contrast to using an SC where an individual is likely to use it infrequently, that is, when they have a medical need.

This study's trust scale was based on Hoffman's Explanation Satisfaction Scale, with the scale tailored toward a layperson user [18]. Four categories of measurement were developed: faith (in the system), reliance (propensity to perform an action based on the explanation), satisfaction (attitude toward the explanation), and comprehension (of the explanation). In all, 3 questions were asked per category. The questions were inspired by scales developed by the human-computer interaction community [80-83], which had many commonalities. The full set of survey questions are presented in Tables S1 and S2 of Multimedia Appendix 2.

## Data Collection

### Participant Recruitment

Participants were recruited using the web-based platform Prolific. Qualtrics (Qualtrics, LLC) was subsequently used to randomly allocate the participants to a treatment group and gather survey responses. Participants who took less than 210 seconds to complete the experiment were removed as this would imply that they took less than 30 seconds to complete the survey, indicating satisficing. The final data set comprised 750 participants who were paid £1.45 (US $1.96 minimum wage equivalent) for their time.

### Ethical Considerations

A number of ethical issues were considered in designing the study, including the following:

- Tests on individuals who are symptomatic require appropriate clinical trial procedures to be followed.
- If participants are not supervised by a clinician, an SC that is 100% accurate must be provided to avoid a misdiagnosis. Such an SC does not currently exist [5].
- Commercial SCs are designed to diagnose numerous diseases, resulting in a plethora of diagnostic pathways. SC model mechanisms are often nondeterministic, making real SCs uncontrollable in an experimental setting.
- The stimuli shown to participants must be medically safe and accurate to avoid misleading them.
- The act of observation could cause the development of health anxiety [84].

To mitigate each of these concerns, nonsymptomatic participants were asked to watch a video of an SC interaction. They were informed that this was not tailored to their own health needs. A General Medical Council–registered primary care physician experienced in SC design advised on the presented materials to ensure medical accuracy, safety, and realism. The participants were limited to individuals residing in the United Kingdom, which offers universal health care, and they were provided with resources to access clinical support if they became concerned. The study received ethical approval (reference: SSHOIIC1A19007) from the University of Oxford.

## Results

### Participants

Data from the 750 participants were analyzed using R (The R Foundation). The participants' ages ranged from 18 to 87 years (mean 35.8, SD 12.6). Of the 750 participants, 512 (68.3%) were in full-time or part-time employment. Most (723/750, 96.5%) had experienced a headache. The majority of respondents in the migraine treatment group viewed their disease as benign compared with a minority of those assigned to temporal arteritis (Table 2).

Table 2. Perceived seriousness of disease by percentage of respondents (N=750; migraine, n=367; temporal arteritis, n=383).

| Seriousness | Disease | |
|---|---|---|
| | Migraine, n (%) | Temporal arteritis, n (%) |
| Not very serious | 256 (69.8) | 40 (10.4) |
| Moderately serious | 105 (28.6) | 266 (69.5) |
| Very serious | 6 (1.6) | 77 (20.1) |

### Exploratory Factor Analysis

The data were subsetted by topic, and exploratory factor analysis was performed to generate dependent variables from the 12 survey questions that measured trust. To test the impact of explanation type on layperson trust (RQ 1), the data were subsetted by disease, allowing the assessment of the impact of varying explanation types while holding disease constant. Similarly, to assess whether knowledge of a disease affects trust in explanation (RQ 2), the data were subsetted by explanation type, allowing the assessment of varying diseases while holding explanation type constant. As the measurement scale quantified different aspects of trust, the underlying factors could be correlated. To allow for correlation, oblimin (oblique) rotation was used.

Depending on the subset, two or three dependent variables emerged. In cases where the loadings resulted in two variables,

these were consistently interpreted as attitudes of *Faith* and *Comprehension*. In cases where the loadings produced three variables, themes of *Faith* and Comprehension still emerged, alongside an additional *Depth* variable. In this context, *Faith* was defined as blind trust in the explanation itself, *Comprehension* as an understanding of the information provided, and *Depth* as the richness of information provided.

For loadings of trust by disease, see Table 3, and for loadings of trust by explanation type, see Table 4.

**Table 3.** Summary of items, factor loadings, and correlations by disease with oblimin rotation where factors are displayed with their interpreted variable names (N=750; migraine, n=367; temporal arteritis, n=383)[a].

| Question | Temporal arteritis | | Migraine | | |
| --- | --- | --- | --- | --- | --- |
| | Faith | Comprehension | Faith | Comprehension | Depth |
| 16[b] | 0.83 | N/A[c] | 0.80 | N/A | N/A |
| 17 | 0.83 | N/A | 0.79 | N/A | N/A |
| 18 | 0.74 | N/A | 0.68 | N/A | N/A |
| 19 | 0.58 | N/A | 0.62 | N/A | N/A |
| 20 | 0.54 | N/A | 0.57 | N/A | N/A |
| 21 | 0.78 | N/A | 0.77 | N/A | N/A |
| 22 | 0.75 | N/A | 0.74 | N/A | N/A |
| 23 | 0.55 | N/A | N/A | N/A | 0.57 |
| 24 | —[d] | —[d] | N/A | N/A | 0.92 |
| 25 | N/A | 0.88 | N/A | 0.83 | N/A |
| 26 | N/A | 0.71 | N/A | 0.65 | N/A |
| 27 | N/A | 0.46 | N/A | 0.42 | N/A |
| 28 | N/A | 0.79 | N/A | 0.76 | N/A |
| **Factor correlations** | | | | | |
| Faith | —[e] | N/A | —[e] | N/A | N/A |
| Comprehension | 0.58 | —[e] | 0.50 | —[e] | N/A |
| Depth | —[d] | —[d] | 0.52 | 0.39 | —[e] |

[a]For clarity, only factor loadings>0.4 are presented.
[b]Survey questions are presented in Table S1 of Multimedia Appendix 2.
[c]N/A: not applicable.
[d]Loadings of questions that were removed and correlations of factors not generated by analysis.
[e]Self-correlations.

**Table 4.** Summary of items, factor loadings[a], and correlations by explanation type with oblimin rotation where factors are displayed with their interpreted variable names (N=750; input influence, n=189; social proof, n=183; counterfactual explanation, n=192; no explanation, n=186).

| >Question | Input influence | | Social proof | | | Counterfactual explanation | | No explanation | | |
|---|---|---|---|---|---|---|---|---|---|---|
| | F[b] | C[c] | F | C | D[d] | F | C | F | C | D |
| 16 | 0.88 | N/A[e] | 0.77 | N/A | N/A | 0.81 | N/A | 0.76 | N/A | N/A |
| 17 | 0.75 | N/A | 0.65 | N/A | N/A | 0.86 | N/A | 0.82 | N/A | N/A |
| 18 | 0.63 | N/A | 0.71 | N/A | N/A | 0.68 | N/A | 0.68 | N/A | N/A |
| 19 | 0.64 | N/A | 0.59 | N/A | N/A | 0.46 | N/A | 0.60 | N/A | N/A |
| 20 | 0.44 | N/A | 0.65 | N/A | N/A | 0.56 | N/A | 0.66 | N/A | N/A |
| 21 | 0.79 | N/A | 0.80 | N/A | N/A | 0.82 | N/A | 0.78 | N/A | N/A |
| 22 | 0.71 | N/A | 0.70 | N/A | N/A | 0.71 | N/A | 0.83 | N/A | N/A |
| 23 | __[f] | __[f] | N/A | N/A | 0.65 | 0.43 | N/A | N/A | N/A | 0.66 |
| 24 | 0.43 | N/A | N/A | N/A | 0.71 | 0.46 | N/A | N/A | N/A | 0.90 |
| 25 | N/A | 0.90 | N/A | 0.75 | N/A | N/A | 0.85 | N/A | 0.86 | N/A |
| 26 | N/A | 0.69 | N/A | 0.60 | N/A | N/A | 0.73 | N/A | 0.70 | N/A |
| 27 | N/A | 0.50 | 0.41 | N/A | N/A | N/A | 0.44 | N/A | 0.59 | N/A |
| 28 | N/A | 0.78 | N/A | 0.79 | N/A | N/A | 0.74 | N/A | 0.86 | N/A |
| **Factor correlations** | | | | | | | | | | |
| F | __[g] | N/A | __[g] | N/A | N/A | __[g] | N/A | __[g] | N/A | N/A |
| C | 0.54 | __[g] | 0.47 | __[g] | N/A | 0.51 | __[g] | 0.50 | __[g] | N/A |
| D | __[f] | __[f] | 0.47 | 0.35 | __[g] | __[f] | __[f] | 0.59 | 0.45 | __[g] |

[a]For clarity, only factor loadings >0.4 are presented.

[b]F: faith.

[c]C: comprehension.

[d]D: depth.

[e]N/A: not applicable.

[f]Loadings of questions that were removed and correlations of factors not generated by analysis.

[g]Self-correlations.

## Comparison of Explanation Trust by Varying Explanation Type

Variances in trust sentiment by explanation type were examined by performing multivariate ANOVAs (MANOVAs) on each disease subset. The MANOVA tests were followed by separate ANOVAs performed on each dependent variable.

For temporal arteritis, MANOVA suggested a marginal effect of explanation type on explanation trust, $V=0.0289$; $F_{6758}=1.85$; $P=.09$. ANOVAs revealed nonsignificant treatment effects on *Faith*, $F_{3379}=1.32$; $P=.27$ and *Comprehension*, $F_{3379}=2$; $P=.11$.

For migraine, MANOVA revealed no significant effect of explanation type on explanation trust, $V=0.0187$; $F_{91,089}=0.759$; $P=.65$. ANOVAs revealed nonsignificant effects on *Faith*, $F_{3363}=0.7$; $P=.55$; *Comprehension*, $F_{3363}=1.13$; $P=.34$; and *Depth*, $F_{3363}=1.34$; $P=.26$.

## Comparison of Explanation Trust by Varying Disease

To investigate varying the disease presented, MANOVAs were performed on each explanation type subset using a single independent variable: the disease provided. MANOVA tests were followed by two-tailed *t* tests performed on each dependent variable.

Input influence ($P=.001$), social proof ($P=.049$), and no explanation ($P=.006$) were found to be significant, with counterfactual explanation ($P=.053$); *t* tests were used as a post hoc test. Means and SDs are presented in Table 5, and the parametric test results are reported in Multimedia Appendix 3.

**Table 5.** Mean scores and SDs for trust in explanation by varying disease presented (N=750; input influence, n=189; social proof, n=183; counterfactual explanation, n=192; no explanation, n=186).

| Explanation type and disease | Faith, mean (SD) | Comprehension, mean (SD) | Depth, mean (SD) |
| --- | --- | --- | --- |
| **Input influence** | | | |
|     Migraine | –0.141 (1.12) | 0.0799 (0.968) | N/A[a] |
|     Temporal arteritis | 0.137 (0.736) | –0.0774 (0.902) | N/A |
| **Social proof** | | | |
|     Migraine | –0.116 (0.931) | 0.0661 (0.883) | –0.0614 (0.939) |
|     Temporal arteritis | 0.113 (0.955) | –0.0639 (0.917) | 0.0594 (0.826) |
| **Counterfactual explanation** | | | |
|     Migraine | 0.023 (0.856) | 0.144 (0.839) | N/A |
|     Temporal arteritis | –0.022 (1.042) | –0.142 (0.984) | N/A |
| **No explanation** | | | |
|     Migraine | 0.0381 (0.974) | 0.219 (0.823) | 0.0134 (0.913) |
|     Temporal arteritis | –0.035 (0.946) | –0.201 (1) | –0.0123 (0.936) |

[a]N/A: not applicable.

### Inclination to Use the SC

An overview of situations where participants would consider using an SC of this nature is presented in Table 6. Chi-square tests revealed no significant differences in these responses. For brevity, statistics are not reported here.

**Table 6.** Inclination to use symptom checkers (N=750; respondents were permitted multiple answers).

| When would you use this type of symptom checker? | Respondents, n (%) |
| --- | --- |
| I would never use this kind of symptom checker | 64 (8.5) |
| Any time I felt poorly | 126 (16.8) |
| If I felt moderately unwell | 317 (42.3) |
| If I couldn't speak to a human clinician | 328 (43.7) |
| In situations where I would currently Google my symptoms | 383 (51.1) |

## Discussion

### Overview

Relatively little is known about the effect of explanations on layperson users' trust in increasingly ubiquitous AI SCs. This study aims to investigate the impact on trust among laypeople of *how*-explanations, which are commonly used by system builders, alongside a theoretically grounded *why*-explanation.

The findings reveal nuanced effects of the explanation that are first examined through varying the explanation type (see *Impact of Varying Explanation Type* section) and subsequently through varying the disease (see *Impact of Varying Disease* section). Finally, the participants' high propensity to use SCs is discussed, indicating that suboptimal explanations may be constraining SCs' ability to deliver health care more effectively (see *Propensity to Use SCs* section).

### Impact of Varying Explanation Type

#### Factor Analysis as an Indication of Trust and Knowledge

As exploratory factor analysis generated different components per disease, the participants seemed to have differing conceptualizations of trust. The clean loading of the migraine responses into three components suggests greater nuance in interpretation compared with the less clear factor structure of temporal arteritis. Temporal arteritis did not generate a *Depth* component, and factor analysis required the removal of a question related to *sufficient detail*. As indicated previously, temporal arteritis was selected with the expectation that recipients generally knew less about temporal arteritis than about migraine.

Given the link between explanation and understanding [37,85], it follows that a recipient's perception of explanation quality gives insight into their mental model of the explanandum. The lack of a clear perception of *Depth*, coupled with the unclear factor loadings, suggests that the participants had less knowledge of temporal arteritis. In other words, there was a larger information gap between participants' general medical knowledge of temporal arteritis compared with their knowledge of migraine. Where an information gap occurred, an asymmetric knowledge condition was created where the participants had a need for knowledge to be transmitted [63].

### Impact of Explanation Type in Trust of Migraine Explanation

Varying the type of explanation provided for migraine resulted in no significant difference in explanatory trust. This is likely attributable to the participants' knowledge of migraine.

Studies have demonstrated that patients evaluate medical information against their underlying knowledge base [86-88]. It is likely that as the participants watched the symptoms being described in the experiment, they relied upon their existing knowledge and formed a hypothesis of what may be the cause. As migraine symptoms are commonly known, the subsequent diagnosis of migraine aligned with the participants' knowledge.

An information gap is strongly predictive of an individual's need to be provided with an explanation [34]. The migraine results are supported by evidence that explanations are only required when there is an information gap between the explanation and existing knowledge [33]. In addition, because migraine is a common disease, it is highly probable; it is known that high priors increase the acceptance of explanations [36]. The lack of effect suggests that for a common, benign, well-known disease, the participants generally did not ask *why* the SC chose the diagnosis or *how* it came to the answer. Either the diagnosis made sense, or it did not. Thus, the explanation is not evaluated [14], and there is no impact of the explanation on the users' trust.

### Impact of Explanation Type on Trust of Temporal Arteritis Explanation

Varying the type of explanation provided for temporal arteritis was reported as marginally significant ($P$=.09). This claim is made because significant findings when examining the effect of varying the disease (see *Impact of Varying Disease* section) add further support to the evidence that an effect is being observed. It is likely that the need for an explanation is stronger for temporal arteritis than for migraine. This explanatory need caused the participants to critically evaluate the temporal arteritis explanation. Before exploring these needs, it is instructive to first consider the stimulus that the participants were shown.

The chatbot video ends by informing the user that the disease requires immediate clinical attention and showing the user booking an appointment with a general practitioner. This severity was noted by the participants, 90% (345/383) of whom judged temporal arteritis to be moderately or very severe (Table 2). When presented with a diagnosis of temporal arteritis, most of the participants would not be knowledgeable of its symptoms. A diagnosis of an unknown, severe disease may have surprised or even alarmed the participants, causing an emotional response.

There are two potential factors that stimulated a greater need for explanation in the temporal arteritis group. First, as a rare disease with lesser-known symptoms, temporal arteritis created an information gap that caused a desire for an explanation (see *Explanation Seeking and Knowledge* section). Second, human emotion is known to affect explanation [37] and can influence how we experience events [89]. Surprise is a known predictor of explanation seeking when evaluating how a stimulus aligns with prior beliefs [34,90]. It is therefore possible that emotions such as surprise or even fear generated by the diagnosis may have provoked a greater need for an explanation. As the measurement of emotion was beyond the scope of this study, this would be a promising direction for future research.

Returning to the main discussion regarding the participants' need for an explanation, the marginal results suggest that an explanation may have improved trust, but there was no clear effect. The lack of significance was surprising, given the presence of a no explanation experimental treatment. Considering the result in the context of the experimental format illuminates possibilities as to why this should be so.

All experimental treatments, even no explanation, involved viewing a 3-minute chatbot interaction. The conversational nature of the interaction transmitted explanatory information [22]. The participants would have consulted their existing medical knowledge to build a hypothesis of the diagnosis while viewing the video. Finally, the description of the disease itself contained factual information. Therefore, the participants likely viewed all treatments, including no explanation, as a form of explanation. This is despite the no explanation type providing no explicit detail of *how* or *why* the SC derived its diagnosis.

The three remaining explanation types provided an answer to a *why* or *how* question. When framed by the need to close the temporal arteritis information gap, it seems that these explanations did not fully succeed. It seems that the explanation was not complete enough to be persuasive [45]. Lack of completeness in a medical explanation highlights a serious issue. There are tens of thousands of illnesses and conditions that can affect humans [91]. Of these, a layperson likely knows a handful of common illnesses and those that are the subject of public awareness campaigns. To a layperson, the explanations presented likely did not rule out other diseases that they were aware of. For example, when a layperson evaluated the symptoms presented by the counterfactual explanation, might they have wondered if headache and jaw pain also indicated a stroke? Returning to facts and foils (see *Everyday Explanations* section), it is logical that individuals can only generate foils of diseases that they are aware of. This is important because currently SCs have no awareness of the specific foil each user generates.

It is possible that the SC explanations are viewed as less effective because of a phenomenon known as causal discounting: "The role of a given cause in producing a given effect is discounted if other plausible causes are present [92]." In this scenario, the cause (temporal arteritis) is discounted in producing the effects (health symptoms) because other diseases could plausibly cause them.

Further support for causal discounting can be found in the text comments left by the participants upon survey completion. Of the 750 participants, 195 (26%) left comments. Of these 195, 59 (30.3%) indicated that they were fearful that a more serious disease may have been missed, equally distributed between the two diseases. This is a common concern, with 55% of patients fearing misdiagnosis [93]. As such, layperson need for reassurance that other serious diseases have been considered and dismissed is an important factor when explaining diagnoses. Although these comments were equally distributed between the two diseases, the need to address causal discounting was greater for the disease that created the larger information gap.

It is possible that the simple explanations provided in this study were not sufficiently complete to close the recipient's information gap [46] and answer their foil. Although humans may prefer simple explanations [44], they do not blindly prefer them, instead calibrating their preferences according to the context of the explanandum [47]. This study's findings support those in previous works where explanations that are viewed as incomplete are less satisfying to the recipient [30,48]. In the case of the complex nuances of a diagnosis, this study aligns with the findings that people expect a level of complexity that matches that of the event itself [48]. This is orthogonal to the conversational maxim by Grice [94] that explanations should not suggest causes they think the explainee knows. In an SC scenario, this study suggests that the SC must explain that it has considered diseases that the explainee is aware of.

Having evaluated the effect of varying the explanation type on trust in an explanation for a given disease, this discussion now moves to the second perspective of this study: holding the explanation type constant and varying the disease being explained.

## Impact of Varying Disease

### Overview

This section builds on the previous finding that the disease being explained is a key factor in determining trust in explanation. As the findings are nuanced, this section sequentially examines the results of each explanation type. Two general themes emerged from this line of enquiry: (1) the uniqueness of symptoms better closes information gaps, and (2) different explanation types prime users to respond according to the explanation's emphasis. In cases where the emphasis is at odds with the user's own mental model, this causes cognitive dissonance, highlighting the danger of providing explanations when a user's explicit foil is not understood.

Before entering this exploration, it should be noted that although the MANOVAs conducted on three of the four explanation types were significant, post hoc $t$ tests were not necessarily so (Table S2 of Multimedia Appendix 2). This suggests that the facets of trust in an explanation combine to create effects. It is also acknowledged that by dividing the respondents into the four explanation type groups, the post hoc tests were underpowered, and larger sample sizes were required for validation.

### Input Influence

The most interesting finding when examining results by explanation type are that with regard to input influence, *Faith* is significantly different between the diseases ($P$=.053), with the temporal arteritis explanation more trusted than the migraine explanation (Table 5). The explanation given for temporal arteritis states a headache with scalp and jaw pain whereas migraine gives slow onset headache and nausea (Table 1). Returning to the discussion of the simplicity-complexity paradox and causal discounting, the distinctiveness of the combined symptoms given for temporal arteritis likely better closed the participants' information gap [95]. These unusual symptoms provide a simple explanation that excludes other diseases that the explainee may be aware of, cultivating greater faith [46]. Meanwhile, headache and nausea are symptoms that could be explained by diseases other than migraine, rendering it less trustworthy. As *Faith* explains the largest amount of variance in the factors of trust, the significance here underscores the marginal result when examining trust by explanation types for temporal arteritis.

Viewing input influence through this lens raises another important point. It is impossible to tell whether the participants perceived the explanation to be a *how*-explanation or a *why*-explanation. The aforementioned reasoning indicates that the participants were making a contrastive evaluation (indicating *why*); yet, we know that an expansion of the input influence explanation would result in a *how*-explanation (see *How-Explanations in AI* section).

### Social Proof

Social proof is the only explanation format, which is clearly answering *how* the SC generated a diagnosis. It is known that the provision of an explanation can influence the importance of different features in understanding category membership [96,97]. In addition, laypeople observing AI systems are known to construct an internal mental model of the cognitive process of the software itself [98,99]. Interestingly, question 27—"It's easy to follow what the system does."—loaded into *Faith*, whereas for other explanation types, this question loaded into *Comprehension* (Table 4). This suggests that the *how* cue of social proof changed the participants' perception of the SC's cognitive mechanism. The participants were implicitly trusting social proof as a clustering technique as opposed to understanding the mechanism itself.

It is surprising that social proof generated a *Depth* factor, given the nature of this explanation. Return to the conversational nature of the SC causing participants to form a mental model of the disease as they observe questions being answered. By not providing an explanation which contains medical information, the recipient is left to evaluate their mental model against the diagnosis. This raises the question: does providing some form of explanation create more questions than it answers for laypeople?

### Counterfactual Explanation

*Comprehension* was found to be significant ($P$=.03), with a small effect size and migraine being better comprehended than temporal arteritis (Table 5). Again, this points to the previous debate around information gaps and existing medical knowledge. The counterfactual explanation for migraine *undid* nausea, a well-known migraine symptom, matching general knowledge of migraines. For temporal arteritis, suggesting "If your scalp and jaw didn't hurt, I'd have suggested you had a migraine" did not sufficiently close the information gap for the participants. In the example, the user *did* have these symptoms. Erasing their existence did not rule out other diseases that the participants were aware of. It is not inconceivable that the participants saw the counterfactual explanation and thought, "But they do have scalp and jaw pain, couldn't these be caused by another serious condition like a stroke?"

Although counterfactual explanations may be theoretically advocated by social science as being effective in SC-type situations, a good explanation must be "relevant to both the question and to the mental model of the explainee [23,95,100]." The counterfactual explanations provided in this study, although medically valid, did not address the participants' *why* questions. This led to a significant impact on comprehension between the diseases; yet, there was no significant impact when comparing explanation types for the same disease. Existing literature calls for the use of counterfactual explanations [38], typically oriented toward expert users. As information gaps are smaller for expert users than for laypeople, the level of information required for transmission may differ. For example, a medical physician would be capable of ruling out serious, common diseases when observing the SC video, whereas the study findings indicate that a layperson may not be able to do so. Again, this stresses the importance of understanding a layperson's specific foil before generating an explanation.

### *No Explanation*

With regard to the no explanation type, *Comprehension* was significant ($P=.002$), with a medium effect size and migraine being comprehended better than temporal arteritis. Notably, the factors generated for no explanation explained approximately 10% more of the variance in data than the factors for other explanation types (Multimedia Appendix 2). The no explanation format did not explain *how* the SC worked or *why* it suggested a particular disease. Again, the participants were left to consult their own medical knowledge. These results build on the discussion that providing an explanation may provoke doubt in the recipient by misaligning with their own mental model of both medical knowledge and of how the SC derives results (see *Social Proof* section).

## Propensity to Use SCs

Despite middling levels of trust in the SC explanations, 91.5% (686/750) of the participants stated that they would consider using this type of SC (Table 6). This demonstrates that a large proportion of the digitally literate lay population would be prepared to use an SC when in real health need.

Of the 367 participants in the migraine group, 211 (57.5%) agreed or strongly agreed that the patient should follow the medical advice provided, as did 71.2% (273/383) of those shown temporal arteritis. It seems that a large proportion of the participants felt that a different intervention was necessary. It is not possible to tell whether the participants who were shown temporal arteritis and felt that the patient should not see a general practitioner urgently believed that the symptoms were benign (and that the patient should stay at home) or so serious that an ambulance should be called. The participants' disagreement with the triage instruction could be due to a concern that other diseases that they believe possible had not been considered and discounted, requiring further medical investigation. It could also be that they perceived that there was a greater emergency. This highlights that SC diagnoses are not subject to automation bias [101], the blind belief in a computer instruction because it has been generated by an intelligent system. Instead, users are skeptical of the SC's results. Returning to the societal principle of using SCs to reduce the burden on health care systems, this study shows that SC explanations must be improved to avoid a second human clinical opinion at an inappropriate triage level, increasing the burden on the health system.

## Limitations

As this study was conducted to mirror the experience of a current SC, there are many confounding factors that could have affected the results. The explanations were examined in isolation; however, in a real-life scenario, intrinsic inputs such as pain, concern, and cognitive impairment may change layperson preference. It is also possible that the sample size was insufficient to detect small effects between the explanation types, given other confounding factors such as trust in the technology itself.

Ultimately, these results suggest that the field is ripe for further exploration. Productive lines for future inquiry could include measuring desire for explanation by disease, assessing information gaps and explanation seeking behaviors pre- and post explanation, emotion engendered by diagnosis and subsequent reaction to explanation, the level of complexity preferred in an SC diagnostic explanation, and understanding of diseases that users are concerned about. Crucially, future research must seek to understand users' foils in all these scenarios.

## Conclusions

Millions of people around the world today are being encouraged to use SCs as a first port of call when seeking nonemergency medical care. Despite the prevalence of SCs, no specific research has been conducted into the effectiveness of delivering AI explanations to laypeople in the SC setting. SCs today provide explanations that explain *how* the AI cognitively derived its decision, although social science literature suggests humans prefer *why*-explanations. High-quality explanations are necessary to engender trust in an AI system, which is particularly important for SCs because lack of trust would cause users to seek second opinions from human clinicians. This additional demand could overburden health care systems and fail to deliver the promoted benefits of using SCs.

Our results suggest that the disease being explained is a primary factor in determining trust in subsequent explanation (RQ 2). This supports the view that when laypeople are presented with realistic scenarios, they use prior knowledge [102]. The disease diagnosed may or may not produce an information gap, resulting in differing needs for explanations and knowledge transmission. Our results show that an SC must explain the explanandum, as well as demonstrate that it has considered diseases that the explainee is aware of, to sufficiently close their information gap. Hence, transmitting knowledge is not as simple as selecting the pertinent cognitive elements of the model. Prior studies on medical diagnosis have abstracted away from human scenarios to isolate explanatory effects [13,30]. This study aligns with the findings that everyday, human explanations are far more nuanced and complex than laboratory-based experiments have suggested [48].

This study also provides some evidence that varying the explanation type affects layperson trust of the explanation (RQ

1), although these results are nuanced. For the well-known disease, information gaps were not cultivated; therefore, varying the explanation type had no effect. For the lesser-known disease, varying the explanation type resulted in marginally significant differences in trust. When the explanation type was held constant and the disease varied, three of the four explanations resulted in significant MANOVAs, indicating that facets of trust interact to create an overall perception of trust. Crucially, post hoc *t* tests, where significant, revealed that for some explanation types, temporal arteritis explanations were more trusted, and for some, migraine explanations were more trusted. The explanation types likely primed the participants to respond to particular signals highlighted within the explanations, resulting in these differences. Despite this, the explanation type did affect the level of trust in the explanation itself. These findings highlight a particular challenge of this study's design: by not knowing a participant's specific foil, the generically constructed explanations could not communicate sufficient knowledge, hampering the evaluation of the *how*-explanation and *why*-explanation formats.

The core finding of this study highlights that in order to close a user's information gap, the AI explanation must be generated with an understanding of that user's unique foil. System builders must not presume to know what question a layperson is asking of the system. Although system builders today work to elucidate the mechanisms of the AI system in a simple format, it is more important to close the gap between a layperson's general medical knowledge and the disease diagnosed. Part of this process must communicate that other diseases the user is aware of have been considered. Thus, system builders need to progress beyond communicating a simple, generic explanation toward the ability to receive real-time foils from a user.

## Acknowledgments

Special thanks to Dr Keith Grimes, who consulted on the chatbot stimuli for medical validity and safety netting in his capacity as a National Health Service general practitioner. Thanks also to Dr Emily Liquin and Dr M Pacer for their stimulating conversations on explanations and mental models. This study would not have been possible without the support of the University of Oxford. Data collection was supported by grants from the Oxford Internet Institute and Kellogg College, Oxford. The article publication fee was provided by Open Access at the University of Oxford.

## Conflicts of Interest

CW and DB were previously employed by, and hold share options in, Babylon Health, which has developed a symptom checker chatbot. Babylon Health had no input or involvement in this study. DB also declares current employment by, and shares in, Apple, which provides a COVID-19 symptom checker. Apple had no input or involvement in this study. BM currently serves in an advisory capacity for GSK Consumer Healthcare and has previously received reimbursement for conference-related travel from funding provided by DeepMind Technologies Limited.

## Multimedia Appendix 1

Methodology and results of available Symptom Checker survey.
[[DOCX File , 32 KB-Multimedia Appendix 1](#)]

## Multimedia Appendix 2

Survey questions asked to study participants.
[[DOCX File , 21 KB-Multimedia Appendix 2](#)]

## Multimedia Appendix 3

Multivariate and univariate analyses of variance of trust in explanation by varying disease presented.
[[DOCX File , 21 KB-Multimedia Appendix 3](#)]

XSL•FO
RenderX

## Abbreviations

**AI:** artificial intelligence
**ANOVA:** analysis of variance
**MANOVA:** multivariate analysis of variance
**RQ:** research question
**SC:** symptom checker